\documentclass[twocolumn]{aastex62}
\usepackage{natbib}
\begin{document}

    \title{Stellar Kinematics and Environment at $z \sim 0.8 $ in the LEGA-C Survey: Massive, Slow-Rotators are Built First in Overdense Environments}
    \author{Justin Cole}
    \affiliation{Department of Physics and Astronomy and PITT PACC, University of Pittsburgh, Pittsburgh, PA 15260, USA}
    
    \author{Rachel Bezanson}
    \affiliation{Department of Physics and Astronomy and PITT PACC, University of Pittsburgh, Pittsburgh, PA 15260, USA}
    
    \author{Arjen van der Wel}
    \affiliation{Sterrenkundig Observatorium, Universiteit Gent, Krijgslaan 281 S9, B-9000 Gent, Belgium}
    \affiliation{Max-Planck Institut f\"ur Astronomie, K\"onigstuhl 17, D-69117, Heidelberg, Germany}
    
    \author{Eric Bell}
    \affiliation{Department of Astronomy, University of Michigan, Ann Arbor, MI 48109, USA}
    
    \author{Francesco D'Eugenio}
    \affiliation{Research School of Astronomy and Astrophysics, Australian National University, Canberra, ACT 2611, Australia}
    \affiliation{Australian Research Council Centre of Excellence for All-sky Astrophysics (CAASTRO)}
    \affiliation{Sterrenkundig Observatorium, Universiteit Gent, Krijgslaan 281 S9, B-9000 Gent, Belgium}
    
    \author{Marijn Franx}
    \affiliation{Leiden Observatory, Leiden University, P.O.Box 9513, NL-2300 AA Leiden, The Netherlands}
    
    \author{Anna Gallazzi}
    \affiliation{INAF-Osservatorio Astrofisico di Arcetri, Largo Enrico Fermi 5, I-50125 Firenze, Italy}
    
    \author{Josha van Houdt}
    \affiliation{Max-Planck Institut f\"ur Astronomie, K\"onigstuhl 17, D-69117, Heidelberg, Germany}
    
    \author{Adam Muzzin}
    \affiliation{Department of Physics and Astronomy, York University, 4700 Keele St., Toronto, Ontario, Canada, MJ3 1P3}
    
    \author{Camilla Pacifici}
    \affiliation{Space Telescope Science Institute, 3700 San Martin Drive, Baltimore, MD 21218, USA}
    
    \author{Jesse van de Sande}
    \affiliation{Sydney Institute for Astronomy, School of Physics, A28, The University of Sydney, NSW, 2006, Australia}
    \affiliation{ARC Centre of Excellence for All Sky Astrophysics in 3 Dimensions (ASTRO 3D)}
    
    \author{David Sobral}
    \affiliation{Department of Physics, Lancaster University, Lancaster LA1 4YB, UK}
    
    \author{Caroline Straatman}
    \affiliation{Sterrenkundig Observatorium, Universiteit Gent, Krijgslaan 281 S9, B-9000 Gent, Belgium}
    
    \author{Po-Feng Wu}
    \affiliation{Max-Planck Institut f\"ur Astronomie, K\"onigstuhl 17, D-69117, Heidelberg, Germany}
    
    \date{\today}

\begin{abstract}
    In this letter, we investigate the impact of environment on integrated and spatially-resolved stellar kinematics of a sample of massive, quiescent galaxies at intermediate redshift ($0.6<z<1.0$). For this analysis, we combine photometric and spectroscopic parameters from the UltraVISTA and Large Early Galaxy Astrophysics Census (LEGA-C) surveys in the COSMOS field and environmental measurements. We analyze the trends with overdensity (1+$\delta$) on the rotational support of quiescent galaxies and find no universal trends at either fixed mass or fixed stellar velocity dispersion. This is consistent with previous studies of the local Universe; rotational support of massive galaxies depends primarily on stellar mass. We highlight two populations of massive galaxies ($\log\mathrm{M_\star/M_\odot}\geq11$) that deviate from the average mass relation. First, the most massive galaxies in the most under-dense regions ($(1+\delta)\leq1$) exhibit elevated rotational support. Similarly, at the highest masses ($\log\mathrm{M_\star/M_\odot}\geq11.25$) the range in rotational support is significant in all but the densest regions. This corresponds to an increasing slow-rotator fraction such that only galaxies in the densest environments ($(1+\delta)\geq3.5$) are primarily (90$\pm$10\%) slow-rotators.This effect is not seen at fixed velocity dispersion, suggesting minor merging as the driving mechanism: only in the densest regions have the most massive galaxies experienced significant minor merging, building stellar mass and diminishing rotation without significantly affecting the central stellar velocity dispersion. In the local Universe, most massive galaxies are slow-rotators, regardless of environment, suggesting minor merging occurs at later cosmic times $(z\lesssim0.6)$ in all but the most dense environments.
 
\end{abstract}


\section{Introduction} \label{sec:1}

Growing evidence from observations of quiescent, early-type galaxies through cosmic time \citep[e.g.,][]{bezanson:09, vDokkum:10, hilz:12, newman.a:12, hilz:13, newman.s:13} and from hydrodynamic simulations in a cosmological setting \citep[e.g.,][]{naab:09, wellons:15, wellons:16, penoyre:17} suggests the importance of hierarchical assembly via gas-poor, minor merging in building today's elliptical galaxies. Cosmological simulations predict that the growth of elliptical galaxies through minor merging should extend their radial profiles \citep[e.g.,][]{lagos:17, lagos:18} and decrease their rotational support \citep[e.g.,][]{frigo:19}. Additionally, as ellipticals continue to grow in mass and size, their rotational support decreases \citep[e.g.,][]{vdwel:08b, vdwel:14, bezanson:18b}, with the tendency for galaxies to transition from rotation-supported systems to pressure-supported systems \citep[e.g.,][]{cappellari:11b, vdSande:13, naab:14}.

In this model, the ordered motions of stellar orbits are averaged out by a series of mergers through cosmic time, creating a direct connection between merging and rotational or dispersion support. Given this, one would expect to find environmental trends in the rotational support of elliptical galaxies driven by their differing merger histories \citep[][]{cappellari:11b}. However, although rotational support has been shown to correlate strongly with stellar mass \citep[e.g.,][]{cappellari:11a, vdSande:13, vdSande:17, vdSande:19, veale:17, bezanson:18a, greene:18}, ellipticals in the nearby Universe do not appear to have additional environmental dependencies \citep[][]{veale:17, greene:18}. This suggests that the processes responsible for diminishing rotational support in massive, elliptical galaxies do so independently of environment or that those trends have been eroded over time.

If the destruction of rotational support is gradual in elliptical galaxies, observations of galaxies at a much earlier epoch could probe an informative period of this process, providing stronger tests of the extended nature of this evolution. However, these observations are challenging, requiring sufficient depths to measure the resolved stellar kinematics and large enough samples to search for environmental trends that have been previously out of reach. Early studies of the shapes and rotational support of quiescent galaxies much closer to their quenching episodes point towards a picture of kinematic evolution post-quenching, although there may be some tension between kinematic and morphological studies. \citet{holden:09} found no evolution in the projected shapes of early-type galaxies, from $z\sim1$ to $z\sim0$, implying the lack of rotational support evolution between these epochs. Studies of the field population have shown at most mild evolution in the shape distribution below $z\lesssim0.7$ \citep{holden:12, chang:13}, while at $z\geq1$, there is a clear and accelerated evolution of field galaxy projected shapes \citep{vdWel:11,chang:13}. At $z\sim2$, several strongly-lensed, massive galaxies \citep{newman.a:18, toft:17}, show significant rotation and the spatially integrated stellar kinematics of 80 quiescent galaxies \citep{belli:17} also suggest increased rotational support. \citet{bezanson:18a} demonstrated that a sample of $\sim100$ quiescent galaxies from an early release of the Large Early Galaxy Astrophysics Census (LEGA-C) have $\sim94\%$ more rotational support than local elliptical galaxies.

In this letter, we extend the analysis presented by \citet{bezanson:18a} to determine whether the rotational support of quiescent galaxies in LEGA-C exhibits a dependence on environment, in addition to stellar mass. In \S\ref{sec:2}, we describe the LEGA-C sample and auxillary data sets used in our analysis. We analyze the trends in environment and stellar properties on rotational support in \S\ref{sec:3}. In \S\ref{sec:4}, we summarize our findings and discuss conclusions. We assume a standard concordance cosmology throughout this analysis ($H_0$ = 70 km s$^{-1}$, $\Omega_M$ = 0.3, $\Omega_\Lambda$ = 0.7).

\begin{figure*}[h]
    \centering
    \includegraphics[width=1\textwidth]{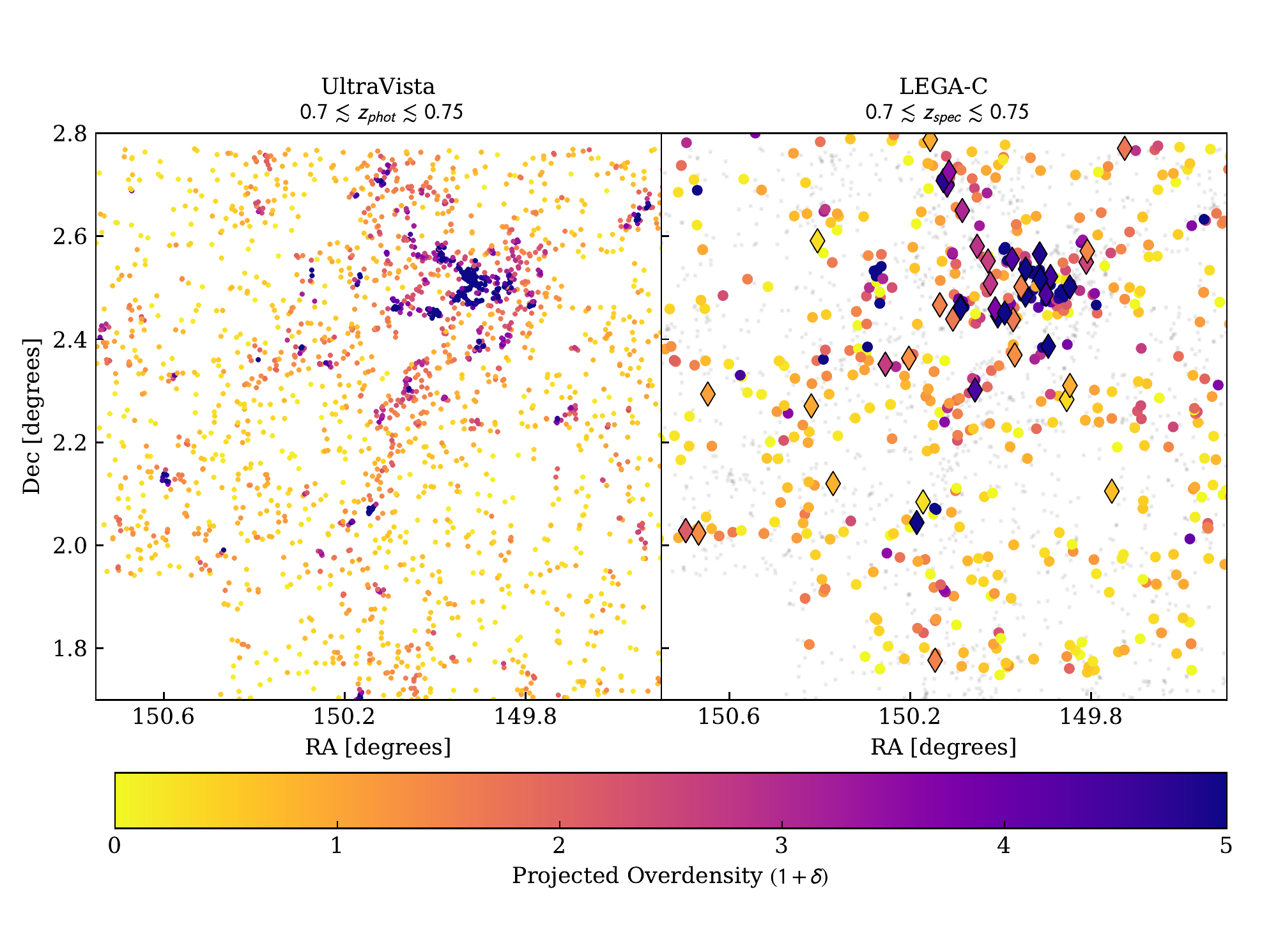}
    \caption{A 2-dimensional projection of the UltraVISTA galaxies (left panel) used in the environmental analysis \citep{darvish:15} and the spectroscopic targets from the LEGA-C survey (right panel) in the COSMOS field for $0.7 < z < 0.75$. Galaxies are colored based on their projected overdensities, from lowest (yellow) to highest (blue). Well-aligned quiescent galaxies used in this analysis are marked as outlined diamonds. LEGA-C targeting sufficiently samples the full range of overdensities in COSMOS.}
    \label{fig:1}
\end{figure*}

\section{Data and Sample} \label{sec:2}

\begin{figure}
    \centering
    \includegraphics[width=0.45\textwidth]{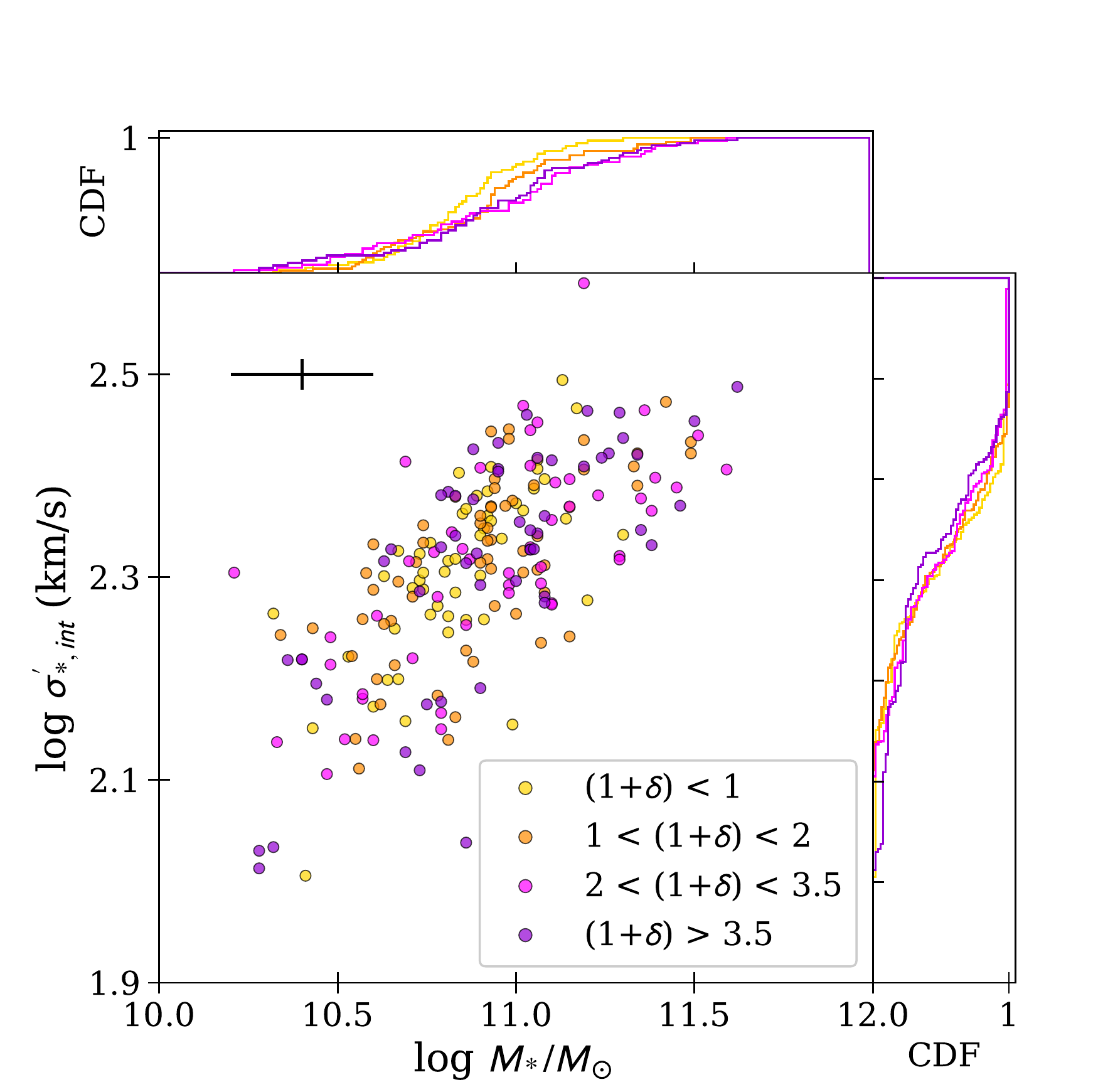}
    \caption{The mass-Faber-Jackson (mFJ) relationship in LEGA-C, colored by overdensity quartiles. We also include the cumulative distribution functions for both stellar mass (top) and $\sigma_{\star, int}'$ (right). The average error for stellar mass and velocity dispersion are shown in the upper left corner of the main panel. Galaxies with higher masses tend to reside in the highest overdensities. Although the trends in $\sigma_{\star,int}$ are more subtle, galaxies residing in the highest overdensities tend to have slightly higher $\sigma'_{\star,int}$.}
    \label{fig:2}
\end{figure}

\subsection{The LEGA-C Spectroscopic Dataset of Massive Galaxies at $z \sim 0.8$}

The sample of galaxies used in this paper is based on LEGA-C data release 2 (DR2) \citep{straatman:18} (PI: van der Wel). LEGA-C includes ultra-deep spectroscopy of approximately 3500 massive galaxies at z $\sim$ 0.8 in the COSMOS field using VIMOS on the VLT as a part of an ESO Large Spectroscopic Program. A more detailed description of the survey, data reduction, and quality can be found in \citet{vdwel:16} and \citet{straatman:18}. Observations were taken using the HRred grating, which produces R $\sim$ 2500 spectra between $\sim$ 6300 and 8800 {\AA}. The LEGA-C survey targets massive galaxies with a redshift-dependent K-magnitude limit (K$_{AB}$ = 20.7 - 7.5 log($\frac{1+z}{1.8}$)) that yields a representative sample of galaxies above $\log M_{\star}/M_\odot \geq 10.4.$ Spectroscopic targets are selected from the \citet{muzzin:13a} v4.1 UltraVISTA catalog, which includes 30 photometric band measurements from 150 {\AA} to 24000 {\AA} from the GALEX, Subaru, Canada-France-Hawaii, VISTA, and Spitzer telescopes. Stellar population properties are estimated for the full sample using FAST \citep{kriek:09} assuming delayed exponentially declining star formation histories, a \citet{chabrier:03} Initial Mass Function, \citet{calzetti:00} dust law and fixing to the spectroscopic redshifts. HST/ACS F814W imaging of each galaxy \citep{koekemoer:07, massey:10} is fit with a S\'ersic profile using \textit{Galfit} \citep{peng:02, peng:10}. We note that all VIMOS slits are North-South aligned in the LEGA-C survey, therefore we restrict our analysis in this work to galaxies for which the photometric major axis is within 30 degrees of the slit.

The spatially resolved stellar kinematics measured from the LEGA-C spectra are vital for this analysis. Full details of the kinematic modelling of the spectra are described in \citet{bezanson:18a, bezanson:18b} and we summarize briefly here. Each 2D and 1D optimally-extracted spectrum is fit using \textit{pPXF} \citep{cappellari:04, cappellari:17} with a non-negative linear combination of theoretical single stellar population templates and Gaussian emission lines and broadened to fit the spectrum. This yields stellar and ionized gas rotation curves and dispersion profiles along the slit for all galaxies in the survey. We draw specific attention to two quantities used in our analysis. $\sigma_{*,int}'$ is the stellar velocity dispersion measured from the spatially integrated, optimally extracted spectrum \citep[see][]{bezanson:18b}. We define rotational support by the ratio between the stellar rotational velocity measured at 5 kpc and the stellar velocity dispersion in the central pixel. To minimize the impact of projection effects, we divide this ratio by $\sqrt{\epsilon/(1-\epsilon)}$ where $\epsilon = 1-b/a$ \citep{bezanson:18a}:
\begin{equation}
    (v_5/\sigma_0)^* = \frac{|v_5|/\sigma_0}{\sqrt{\epsilon/(1-\epsilon)}}
\end{equation}
Systematic differences between this observed quantity and the intrinsic rotational support are very likely functions of mass and $\sigma_{\star,int}$. For our study, we focus on a sample of 217 quiescent galaxies, selected by U-V and V-J colors according to \citet{muzzin:13a}, most of which are visually early-type. We do not expect any uncertainty in rotational support to be a function of environment.

\begin{figure*}[t]
    \centering
    \includegraphics[width = 1\linewidth]{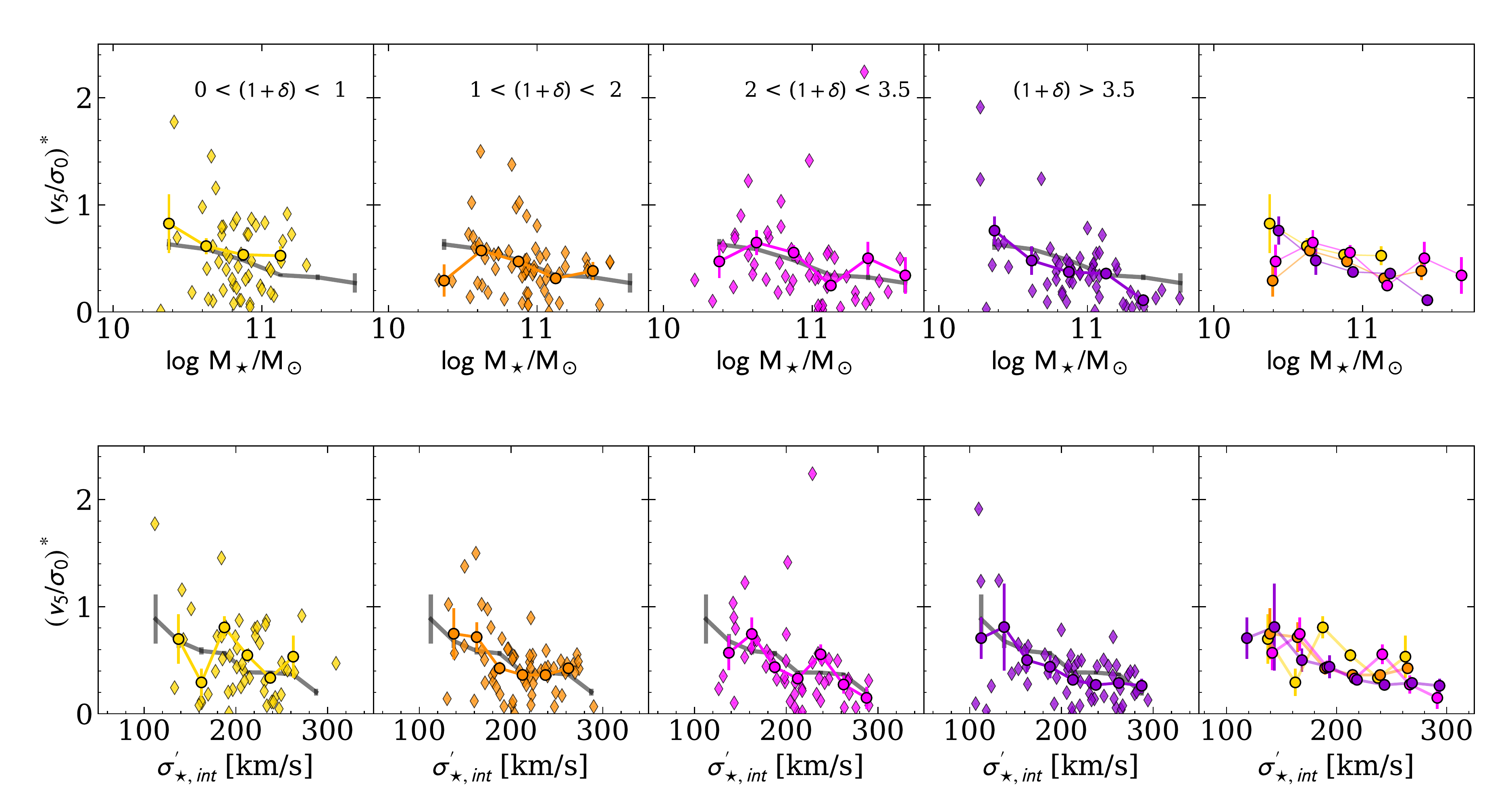}
    \caption{The observed rotational support at 5 kpc, $(v_5/\sigma_0)^*$, of galaxies  versus stellar mass (top row) and $\sigma'_{\star, int}$ (bottom row), binned and colored according to overdensity as in Figure \ref{fig:2}. Black and colored lines show the average rotational support for the entire sample and for each overdensity range, respectively, in mass bins of 0.2 dex with jackknife error estimation. In the two right-most panels, we show the running average rotational support with colored, outlined circles representing the errors in each mass bin, with slight horizontal offsets for clarity. While there is not a clear universal trend, the most massive galaxies in each overdensity quartile exhibit different distributions of rotational support. The most massive galaxies ($11\leq\log\mathrm{M_\star/M_\odot}\leq11.25$) in the least dense environments (yellow, $(1+\delta)\leq1$) have elevated average $(v_5/\sigma_0)^*$. In more dense environments, where the most massive galaxies are larger ($11.25\leq\log\mathrm{M_\star/M_\odot}$), the average $(v_5/\sigma_0)^*$ is significant in all but the densest regions (($1+\delta)>3.5)$.}
    \label{fig:3}
\end{figure*}

\subsection{LEGA-C and Sampling the COSMOS Field}

In addition to the LEGA-C dataset, we include information about galaxy environments in the COSMOS field, focusing on projected overdensities (1+$\delta$) from \citet{darvish:17}. This group catalog uses the COSMOS2015 photometric redshift catalog from \citet{laigle:15} in the UltraVISTA-DR2 region \citep{mccracken:12, ilbert:13}. Adaptive weighted kernel smoothing is used to determine projected number densities and subsequent overdensities. The projected densities are determined using a 2-dimensional Gaussian kernel which changes depending on the local density of galaxies within each redshift slice. For a more complete description, see \citet{darvish:15}. There is an additional component of randomness added in the measurements of overdensity, which we expect to smear out any trends related to environment. We match LEGA-C galaxies to the group catalog within 1". 

Although LEGA-C is a targeted sample, it traces the full range of overdensities. In Figure \ref{fig:1}, we show a 2-dimensional projection of the photometric UltraVISTA galaxies (left) used in the environmental analysis \citep{darvish:15} and a sub-sample of the spectroscopic targets from the LEGA-C survey (right) in COSMOS for a small redshift slice ($0.7 \leq z \leq 0.75$). Galaxies are colored by their projected overdensity (1+$\delta$) and we have marked the well-aligned quiescent galaxies used in this analysis with outlined diamonds. The range in log (1+$\delta$) for the UltraVISTA photometric catalog is $0.01 \leq (1+\delta) \leq 35.36$ and the range sampled by the LEGA-C survey is $0.3 \leq (1+\delta) \leq 21.81$, which effectively spans the full dynamic range of overdensities in the COSMOS field.

\subsection{Nearby quiescent galaxies from the MASSIVE and ATLAS $^{ 3D}$ surveys}

Finally, we include a comparison sample of massive, quiescent galaxies in the local Universe from the MASSIVE and ATLAS$^{3D}$ surveys. The MASSIVE survey is a volume-limited sample of 115 galaxies in which all galaxies with a K-band magnitude brighter than $M_k$ $\leq-23.5$ are targeted \citep{carrick:15} and observed using an integral field (IFU) spectrograph giving 2-dimensional stellar kinematic information about each galaxy \citep{veale:17}. ATLAS$^{3D}$ is also an IFU survey, observing all 260 galaxies above $M_k \leq -21.5$ and within a 42 Mpc radius. For a complete description of the ATLAS$^{3D}$ survey, see \citet{cappellari:11a}. For the purpose of this paper, we use the stellar kinematic parameter $\lambda_\epsilon$ \citep{emsellem:11,veale:17} to quantify rotational support and classify galaxies in the local Universe as fast-/slow-rotators.
Using a linear $M_K$-to-stellar mass ratio \citep{cappellari:13a}, we convert the K-band magnitudes of galaxies in the MASSIVE and ATLAS$^{3D}$ surveys and compare them to galaxies in the LEGA-C survey with the highest masses. There is $\sim 0.3$ dex uncertainty in stellar masses which comes from uncertainties in the K-band magnitudes and the $M_\star-M_K$ relation \citep{cappellari:13a}. $\lambda_\epsilon$ is measured by binning the spatial pixels in each galaxy until a signal-to-noise threshold of 20 is reached, and averaging the bins out to the effective radius of the galaxy. To specify the environment of MASSIVE and ATLAS$^{3D}$, we adopt luminosity-weighted overdensities $(1+\delta_g)$, taken from \citet{carrick:15} and \citet{lavaux:11}, respectively. ATLAS $^{3D}$ and MASSIVE have volumes of $\sim10^{5}$ $\mathrm{Mpc^3}$ and $\sim10^{6}$ $\mathrm{Mpc^3}$, respectively, and LEGA-C has a volume of $\sim3\times10^{5}$ $\mathrm{Mpc^3}$.

\section{Dependence of Rotational Support on Environment at $z \sim 0.8$} \label{sec:3}

\begin{figure*}
    \centering
    \includegraphics[width=1\linewidth]{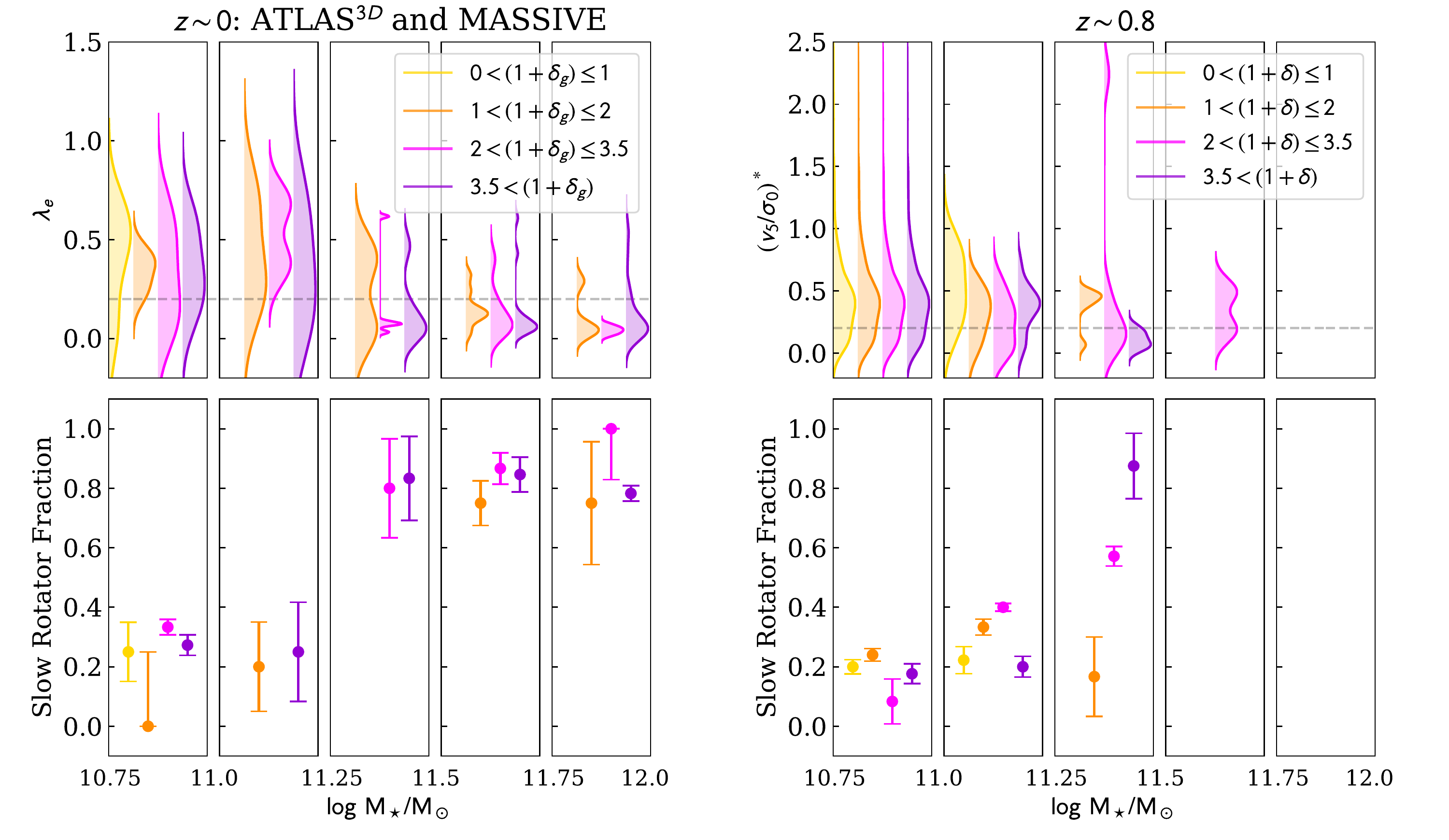}
    \caption{\textit{(Top row:)} The normalized probability distribution functions (PDFs) for $\lambda_\epsilon$ (left, ATLAS $^{3D}$ and MASSIVE) and $(v_5/\sigma_0)^*$ (right, LEGA-C) for the most massive galaxies in both samples ($\log\mathrm{M_\star/M_\odot}\geq10.75$). The dashed horizontal line shows the slow-rotator threshold (rotational support $\leq0.2$). \textit{(Bottom row:)} The slow-rotator fractions for ATLAS $^{3D}$ and MASSIVE (left) and LEGA-C (right). PDFs and points are colored by overdensity. Both LEGA-C and the nearby sample show similar slow-rotator fractions at the lowest masses. However, for galaxies with $\log\mathrm{M_\star/M_\odot}\geq11.25$, there is a clear separation in the slow-rotator fractions for LEGA-C galaxies while those for the nearby Universe only exhibit a trend with mass.}
    \label{fig:4}
\end{figure*}

In this section, we investigate whether stellar kinematics in quiescent galaxies at z $\sim$ 0.8 depend on environment. In the primary panel of Figure \ref{fig:2}, we show the distribution of observed stellar velocity dispersion versus stellar mass, or the mass Faber-Jackson relation \citep[mFJ,][]{faber:76} for all the galaxies in the LEGA-C sample, colored by overdensity, with each bin containing $\sim50$ galaxies. We continue this color scheme in later figures. Cumulative distribution functions (CDFs) are shown for both stellar mass and $\sigma_{\star, int}'$. Stellar mass tends to increase with overdensity, with the most massive galaxies accumulating in the densest environments. However, the trend is more subtle in the CDFs for $\sigma_{\star, int}'$: at most overdensities, galaxies tend to have similar $\sigma_{\star,int}'$ except in the highest overdensities, where galaxies tend to have the highest $\sigma_{\star,int}'$.

Additionally, we investigate trends in rotational support with environment. In Figure \ref{fig:3}, we show the rotational support of galaxies $(v_5/\sigma_0)^*$ versus stellar mass in the top row and versus $\sigma_{\star, int}'$ in the bottom row, colored by overdensity. Black lines and colored lines show the average rotational support for the entire sample and for each overdensity bin, respectively, with jackknife error estimation. In the two right-most panels, we show the running average rotational support for each overdensity bin, with a slight offset from the center of the bin for clarity. As shown in \citet{bezanson:18b}, the average range in rotational support tends to decrease with increasing stellar mass, which is consistent with studies of massive, quiescent galaxies in the local Universe \citep{veale:17, greene:18}. We do not see a strong environmental trend at all masses, but we note two statistically significant trends at the massive end of the sample. First, while galaxies in the least dense environments (yellow symbols) are not represented at the highest masses ($\log\mathrm{M_\star/M_\odot}\geq11.25$), the most massive of these ($11\leq\log\mathrm{M_\star/M_\odot}\leq11.25$) exhibit more rotational support than other similar mass galaxies. In denser environments ($(1+\delta)>1$),  massive galaxies follow the average relation except at the highest masses ($\log\mathrm{M_\star/M_\odot}\geq11.25$), where only galaxies in the most overdense regions have minimal $(v_5/\sigma_0)^*$. Unlike comparisons at fixed mass, trends in $(v_5/\sigma_0)^*$ at fixed $\sigma'_{\star,int}$ are much more subtle.

\begin{figure*}
    \centering
    \includegraphics[width=1\linewidth]{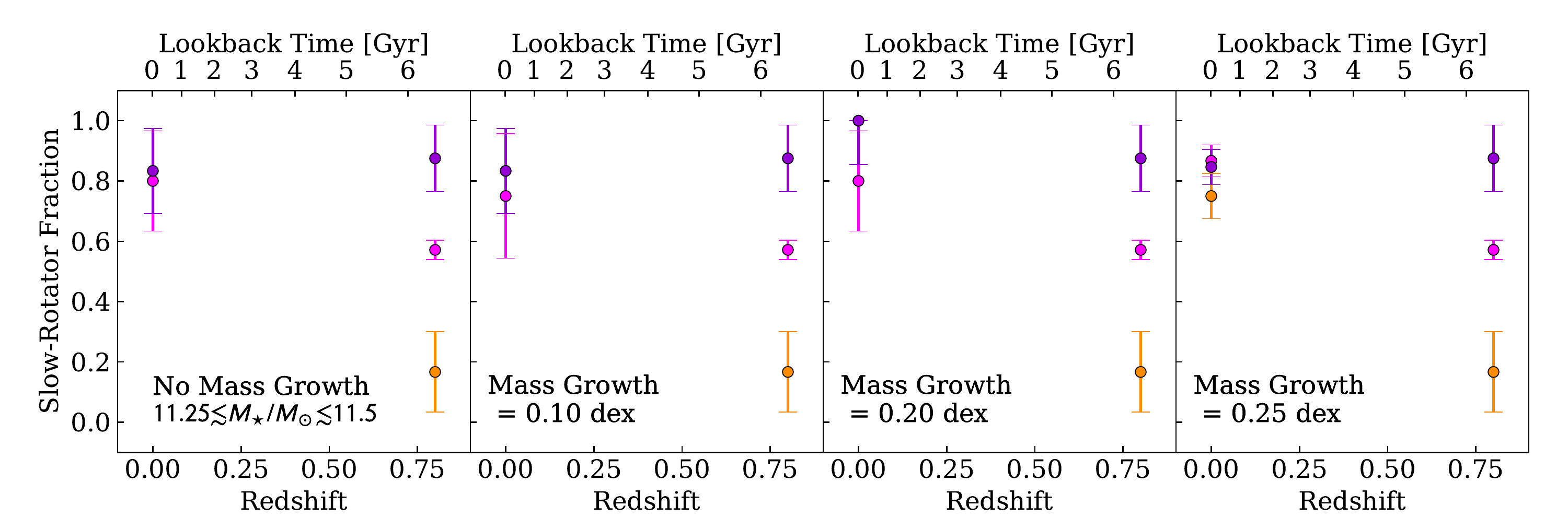}
    \caption{The slow-rotator fractions versus redshift, or lookback time, for massive quiescent galaxies ($11.25\leq\log\mathrm{M_\star/M_\odot}\leq11.5$) at $z\sim0.8$, colored by overdensity. From left to right we assume galaxies grow by 0, 0.10, 0.20, and 0.25 dex in the 6 Gyr span between the surveys. Unlike the trend in high-mass quiescent galaxies in LEGA-C, there is no significant dependence on environment in any possible population of nearby galaxies. This implies that the most massive galaxies in the most overdense regions were kinematically evolved by $z\sim0.8$, but that those residing in lower density regions must undergo significant subsequent evolution, likely driven by minor merging, to resemble any slow-rotating early-type galaxies today.}
    \label{fig:5}
\end{figure*}

We focus the remainder of the letter on the most massive galaxies in the sample ($\log M_\star/M_\odot \geq 11.25$). In the local Universe this corresponds to the mass at which galaxies are primarily slow-rotators, or core ellipticals \citep[e.g.,][]{cappellari:13a, cappellari:13b}. In the top row of Figure \ref{fig:4} we show the Gaussian-kernel smoothed, normalized probability distribution functions \citep[][]{seaborn:v7} for $\lambda_\epsilon$ (left, ATLAS $^{3D}$ and MASSIVE) and $(v_5/\sigma_0)^*$ (right, LEGA-C) for galaxies with log $M_\star/M_\odot\geq10.75$, in bins of 0.25 dex. We note that the different distributions identified in Figure \ref{fig:3} likely correspond to a difference in populations of so-called fast- and slow- rotators in the local Universe. The dashed horizontal line indicates the separation between fast- and slow-rotators. In the bottom row, we indicate the fraction of galaxies in each mass and density bin that lie below the slow-rotator threshold. We adopt a threshold of $\lambda_\epsilon=0.2$ to discriminate between the two populations following \citet{veale:17}; however, using a threshold of $\lambda_\epsilon=0.2\sqrt{\epsilon}$ does not significantly affect the identification of slow-rotators in this sample of massive galaxies. While $(v_5/\sigma_0)^*$ is an empirical quantity and does not have an agreed upon threshold to separate galaxies with significant rotation and those without, we adopt a threshold of 0.2 based on the distribution of galaxies (e.g., in Figure \ref{fig:3}). We have tested additional values for this threshold between 0.1 and 0.3, which do not change the results of this analysis; the distributions of galaxies in the distant and local Universe are fundamentally different. 

In the local Universe, the fraction of slow rotators at fixed mass does not depend on environment. In the distant Universe, in the low-density regions ($(1+\delta)\leq1$) only $\sim20\%$ of the most massive ($10.75\leq\log\mathrm{M_\star/M_\odot}\leq11.25$) galaxies exhibit minimal rotation. Although under-dense regions tend to be populated by galaxies with higher average rotational support, as shown in Figure \ref{fig:4}, this does not correspond to a statistically significant difference in the fraction of slow-rotators. This is not true for the distributions of galaxies in denser regions, which tend to decrease in $(v_5/\sigma_0)^*$ with increasing mass. We note that although $\lambda_\epsilon$, $(v_5/\sigma_0)^*$ and stellar masses are measured very differently in the local and LEGA-C samples, they correspond to qualitatively similar properties. In the local samples, the majority of galaxies with $\log M_\star/M_\odot \geq 11.25$ are slow-rotators. However, for galaxies in LEGA-C, the slow-rotator fraction of the most massive galaxies depends strongly on environment; specifically, in the most overdense regions, nearly all ultra-massive ($\log \mathrm{M_\star/M_\odot} \geq 11.25$) galaxies are slow-rotators, while galaxies in less dense environments are progressively more likely to retain significant stellar rotational support.

Finally, we compare the slow-rotator fractions in possible progenitor and descendant quiescent galaxy populations. In Figure \ref{fig:5}, we show the slow-rotator fraction versus redshift for local and distant galaxies, colored by overdensity. Each panel compares the most massive LEGA-C progenitors to local descendant populations, showing the $z\sim0$ slow-rotator fractions for mass ranges of $11.25\leq{\log}M_\star/M_\odot<11.50$, $11.35\leq{\log}M_\star/M_\odot<11.60$, $11.45\leq{\log}M_\star/M_\odot<11.70$, and $11.50\leq{\log}M_\star/M_\odot<11.75$, (allowing for an increase in mass of 0.0, 0.10, 0.20, and 0.25 dex) from left to right respectively. Empirically motivated work \citep[e.g.,][]{leja:13, patel:13, vDokkum:13} and theoretical studies \citep{behroozi:13, torrey:15, torrey:17} have estimated mass growth rates of ~0.15 dex for massive LEGA-C-like galaxies since $z\sim1$, although this value is particularly uncertain at the massive end. At these masses, all potential descendant populations are dominated by slow-rotators, independent of environment. However, the highest mass galaxies in the distant universe display a clear trend with environment: specifically, those in the densest regions tend to mainly be slow-rotators, with the fraction of slow-rotators decreasing with decreasing overdensity.

\section{Discussion and Conclusions} \label{sec:4}

In this letter we investigate the environmental effects on the stellar kinematics of massive, quiescent galaxies at intermediate redshift. We use two quantities for representing stellar kinematics: (1) $\sigma'_{\star,int}$, the spatially integrated, optimally extracted stellar velocity dispersion and (2) $(v_5/\sigma_0)^*$, the projection-corrected ratio between stellar velocity measured at 5 kpc and stellar velocity dispersion in the central pixel. We also compare the slow-rotator fractions of our sample at intermediate look-back time to the slow-rotator fractions for a sample of galaxies in the local Universe. 

Similar to the trends found in the local Universe, our sample of ETGs demonstrates a strong mass and stellar velocity dispersion dependence, and no universal environmental dependence, in rotational support. Although overdense regions tend to host more massive galaxies, the trends with overdensity in $\sigma'_{\star,int}$ are much more subtle. Specifically, only at the highest $\sigma'_{\star,int}$ is there at subtle separation in the CDFs; $\sigma'_{\star,int}$ increases with increasing overdensity.  However, unlike galaxies at $z\sim0$, at $z\sim0.8$, the most massive population of quiescent galaxies is only dominated by slow-rotators in the most overdense environments. Specifically, in highly populated regions, elliptical galaxies tend to be slow-rotators at both redshifts, however in less dense regions, the fractions of slow-rotators increase dramatically between $z\sim0.8$ and $z\sim0$. In contrast, the vast majority of likely descendants in the local Universe of such massive galaxies (e.g. as probed by the ATLAS$^{3D}$ and MASSIVE surveys) are slow-rotators. We do not find any significant environmental dependence in rotational support of the highest $\sigma'_{\star, int}$ galaxies which is consistent with \citet{vDokkum:10}. In this framework, the continued evolution of galaxies must not significantly change the stellar velocity dispersions of massive galaxies in higher density regions. When taken together, we infer that minor merging is the driving mechanism in building the population of slow-rotating, ultra-massive galaxies in overdense regions of the COSMOS field because it can increase mass and black diminish rotational support without significantly influencing central stellar velocity dispersions \citep[e.g.,][]{bezanson:09, vDokkum:10, newman.a:12, newman.s:13}.

A quantitative analysis of the evolution of the rotational support of quiescent galaxies through cosmic time would require self-consistent analysis of both low- and high-redshift samples. We have limited our comparison to fast and slow-rotator fractions, but directly comparing the rotational support within the two samples would need to take into account differences in observations (e.g., seeing, aperture effects, IFU versus slit spectroscopy) and consistent modeling of the kinematics (e.g., Jeans modeling, van Houdt, et al., in prep). Such analysis may reveal additional environmental trends in the kinematics of massive, quiescent galaxies. 

The strongest test of this evolution as a function of time would ideally probe to even earlier cosmic epochs to observe the formation of these massive galaxies. The James Webb Space Telescope will be equipped with the NIRSpec IFU, which will be able to spatially resolve the light from much more distant progenitors of massive slow rotating galaxies. However, the continuum spectroscopy necessary to probe stellar kinematics will be challenging even for spatially integrated measurements. For individual targets, continuum spectroscopy will be possible, but statistical samples will be out of reach for JWST \citep{newman.a:19}. Thirty meter class telescopes with larger apertures and adaptive optics that enable near diffraction-limited seeing will be able to push spectroscopic observations of massive galaxies to higher redshifts, allowing spatially-resolved spectra to be obtained for higher redshifts than is currently possible and probing new epochs of galaxy formation. 

\acknowledgements
This research made us of Astropy \citep{astropy}. Based on observations collected at the European Organisation for Astronomical Research in the Southern Hemisphere under ESO programme 194.A-2005. JC and RB would like to thank Brett Andrews, Jenny Greene, Brad Holden, Jeff Newman, Alan Pearl, David Setton and Lance Taylor for meaningful conversations that contributed to this project and the Pennsylvania Space Grant Consortium for funding this research. RSB gratefully acknowledges funding for project KA2019-105551 provided by the Robert C. Smith Fund and the Betsy R. Clark Fund of The Pittsburgh Foundation.

\end{document}